\documentclass[twocolumn,prl,showpacs,preprintnumbers,amsmath,amssymb]{revtex4}
\usepackage{graphicx}

\begin{document}

\title{Flux Line Lattice Melting and the Formation of a Coherent 
Quasiparticle Bloch State in the Ultraclean 
URu$_2$Si$_2$ Superconductor}

\author{R. Okazaki,$^{1,\ast}$  Y. Kasahara,$^{1}$ H. Shishido,$^{1}$ 
M. Konczykowski,$^{2}$ K. Behnia,$^{3}$ \\
Y. Haga,$^{4}$ T.~D. Matsuda,$^{4}$ Y. Onuki,$^{4,5}$ 
T. Shibauchi,$^{1,\dag}$ and Y. Matsuda$^{1,6,\ddag}$}
\affiliation{
$^1$Department of Physics, Kyoto University, Kyoto 606-8502, Japan\\
$^2$Laboratoire des Solides Irradi\'es, CNRS-UMR 7642 \& CEA/DSM/DRECAM, Ecole Polytechnique, 91128, Palaiseau, France\\
$^3$Laboratoire de Physique Quantique (CNRS), ESPCI, 10 Rue de Vauquelin, 
75231 Paris, France\\
$^4$Advanced Science Research Center, Japan Atomic Energy Agency, Tokai 319-1195, Japan\\
$^5$Graduate School of Science, Osaka University, Toyonaka, Osaka 560-0043, Japan\\
$^6$Institute for Solid State Physics, University of Tokyo, Kashiwanoha, Kashiwa, Chiba 277-8581, Japan}

\begin{abstract}
We find that in ultraclean heavy-fermion superconductor URu$_2$Si$_2$ ($T_{c0}=1.45$~K) a distinct flux line lattice melting transition with outstanding characters occurs well below the mean-field upper critical fields. We show that a very small number of carriers with heavy mass in this system results in exceptionally large thermal fluctuations even at subkelvin temperatures, which are witnessed by a sizable region of the flux line liquid phase. The uniqueness is further highlighted by an enhancement of the quasiparticle mean free path below the melting transition, implying a possible formation of a quasiparticle Bloch state in the periodic flux line lattice.
\end{abstract}

\pacs{74.25.Fy,74.25.Op,74.25.Qt,74.70.Tx}

\maketitle

URu$_2$Si$_2$ 
is a heavy-fermion superconductor \cite{Pal85,Amitsuka}, which has attracted much attention because of the so-called ``hidden order'' transition at 17.5~K, whose order parameter is still a mystery. In addition, recent studies using extremely clean single crystals having huge $\omega_c\tau$ values (where $\omega_c$ is the cyclotron frequency and $\tau$ is the scattering time) \cite{Ohk99,Kas07} reveal that the superconducting state with low transition temperature ($T_{c0}=1.45$~K) embedded in the hidden order state is also uniquely unusual. According to several experiments \cite{Beh05}, most of the carriers disappear below the hidden order transition, resulting in a very low carrier density ($\sim0.02$ electrons/U-atom), reminiscent of semi-metals such as bismuth and graphite. The electronic structure with such heavy low-density carriers hosts exotic superconducting state, including a first-order transition at the upper critical fields $H_{c2}$, a possible chiral $d$-wave symmetry \cite{Kas07}, and quasiparticle (QP) transport in the quantum limit \cite{Ada07}. 

Here we report that the small number of carriers with heavy mass and with extreme cleanness of URu$_2$Si$_2$ also give rise to an extraordinary vortex state, {\it i.e.} giant thermal fluctuations even at subkelvin temperatures and a possible formation of a coherent QP Bloch state. Owing to the exceptionally few disorder we are able to observe the thermally driven melting transition from flux line (FL) lattice into the FL liquid well below the mean-field transition temperature $T_c(H)$. It is manifest by a sharp resistivity drop under a magnetic field, which is in clear contrast to the case in conventional low-$T_c$ superconductors, where the resistivity drops at $T_c$. Despite the low $T_c$, it is in analogous to the first-order melting transition observed at much higher temperatures in clean high-$T_c$ cuprates \cite{Kwo92,Zel95,Sch97,Wel96}. It is quite remarkable that the melting transition continues to very high fields ($\sim 0.7H_{c2}$), which is in contrast to the case of cuprates where quenched disorder terminates the melting transition at low fields ($\lesssim 0.3H_{c2}$) and causes the disorder-driven glass transition from an ordered (Bragg glass) state to a highly disordered glass state \cite{Pal00,Nis00}. Furthermore, an unusual change in the QP mean free path below the melting transition is observed, indicating that the quasiparticles are scattered less by the FL lattice than liquid, which has never been observed in any other type-II superconductors. This points to a possible formation of a novel Bloch-like state by the periodic FL lattice.

\begin{figure}[t]
\includegraphics[width=8.5cm]{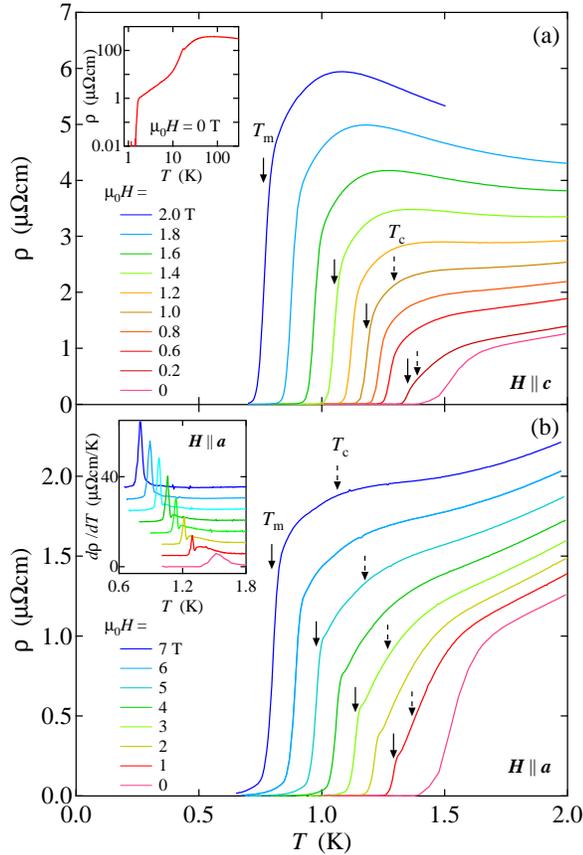}
\caption{(Color online). Temperature dependence of the resistivity for (a) {\boldmath $H$} $\parallel c$ and (b) {\boldmath $H$} $\parallel a$.  The very large magnetoresistance in the normal state stems from the compensation, {\it i.e.} essentially equal number of electrons and holes, $n_e=n_h$ \protect{\cite{Kas07}}. The solid arrows indicate the melting transition $T_m$, which is defined as a peak of $d\rho/dT$ [inset of (b)].  The dashed arrows indicate the mean field transition temperature $T_c$ determined by the cusp of the thermal conductivity [see Fig.~3].    The inset of (a) shows the resistivity in zero field.  The inset of (b) shows $d\rho/dT$ vs. $T$.  The data is vertically shifted.  }
\end{figure}

The ac-resistivity is measured in a $^3$He cryostat by the standard four-probe method with a current density $J$ of 10$^4$~A/m$^2$ along the $a$ axis at 17~Hz.  The electric field $E$ vs $J$ characteristics is analyzed from ac-resistance data with three different excitations.  No evidence for heating of the crystal is obtained down to $\sim 0.7$~K. The thermal conductivity $\kappa$ is measured in a dilution refrigerator by a steady state method along the $a$-axis.

\begin{figure}[t]
\includegraphics[width=9cm]{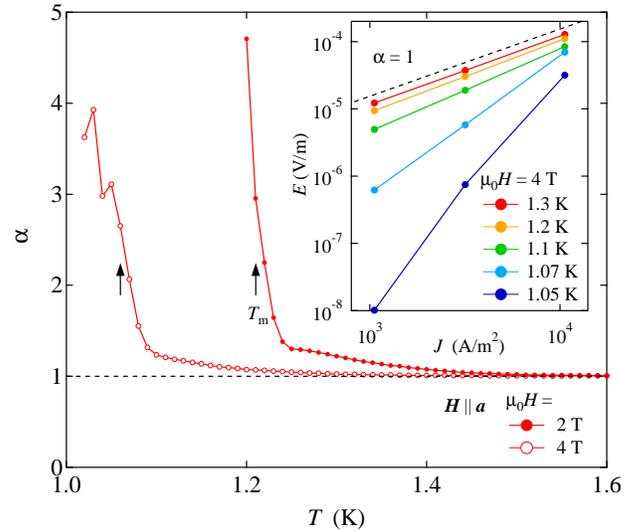}
\caption{(Color online). Exponent $\alpha$ in $E \propto J^{\alpha}$,  as a function of temperature for {\boldmath $H$} $\parallel a$ at 2~T and 4~T.  The inset shows the $E$-$J$ characteristics around $T_m$.  The Ohmic resistivity $\alpha=1$ is observed at high temperatures and $\alpha$ increases gradually with approaching $T_m$ and shows a sharp increase at $T_m$.
}
\end{figure}

In this study, we use a crystal having exceptionally low residual resistivity $\rho_0=0.5~\mu \Omega$~cm and large residual resistivity ratio $RRR=670$ [inset of Fig.~1(a)], which attest the highest crystal quality currently achievable. The resistivity $\rho$ in magnetic fields shown in Fig.~1 exhibits a ``kink" or very sharp drop, which is even sharper than the transition width $\Delta T_{c0}$(10\%-90\%) $\approx0.1$~K in zero field. The kink temperatures determined by the peak positions of $d\rho/dT$ [inset of Fig.~1(b)] are denoted by $T_m$ (solid arrows).   As shown in Fig.~2, the exponent $\alpha$ in the $E$-$J$ characteristics, $E \propto J^{\alpha}$,  exhibits a steep increase near $T_m$ within a narrow temperature range.  

\begin{figure}[t]
\includegraphics[width=9cm]{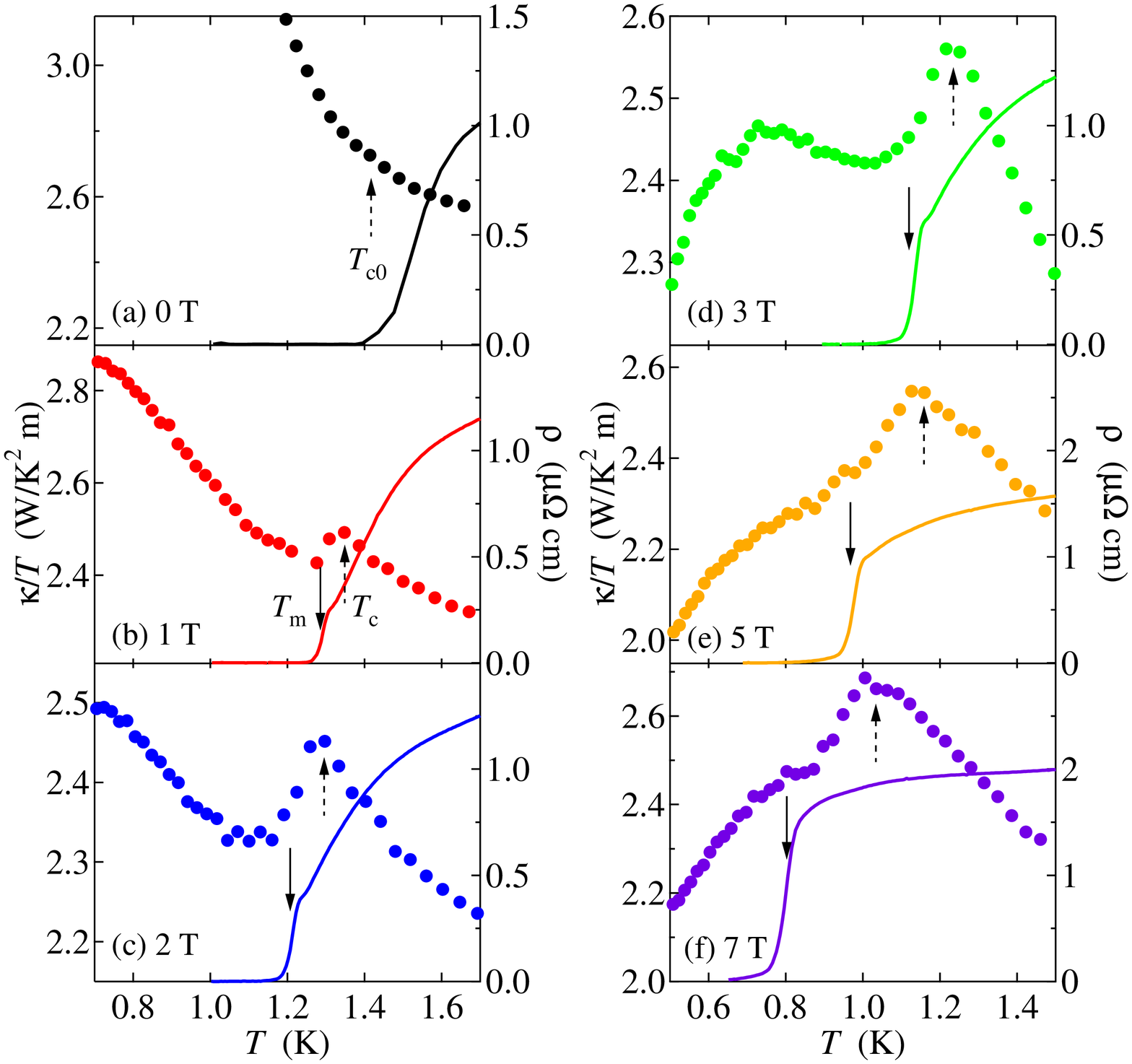}
\caption{(Color online). Temperature dependence of the thermal conductivity divided by temperature $\kappa/T$ (circles) in zero field (a) and in magnetic fields {\boldmath $H$} $\parallel a$ (b)-(f).  The resistivity data at corresponding fields are also shown (solid lines).  The dashed arrows indicate the temperature at which $\kappa/T$ shows cusps, which correspond to the mean-field $T_c(H)$.  The solid arrows mark the melting temperature $T_m$ [inset of Fig.~1(b)].  Below $\sim T_m$, $\kappa/T$ becomes bigger than extrapolated values. 
}
\end{figure}

The thermal conductivity $\kappa$ also provides important information on the vortex states [Fig.~3]. In zero field, $\kappa/T$ increases with decreasing $T$ and more rapidly below $T_{c0}$. 
The Wiedemann-Franz ratio $L$ in the normal state near $T_c$ is very close to the Sommerfeld value $L_0$ expected for the electronic contribution.  Moreover, the phonon contribution $\kappa_{ph}/T$ is reported to be less than 0.3~W/K$^2$m around 0.8~K \cite{Beh92}, which is much smaller than the observed $\kappa/T$. These indicate that in this temperature range, the electron contribution well dominates over the phonon contribution. The electronic heat conduction is described by $\kappa/T\sim N(0)v_F\ell$, where $N(0)$, $v_F$, and $\ell$ are the QP density of states, Fermi velocity, and QP mean free path, respectively.  The enhancement of $\kappa/T$ below $T_{c0}$ is caused by a striking enhancement of $\ell$ due to the gap formation, which overcomes the reduction of $N(0)$ in the superconducting state, as observed in several strongly correlated electron systems \cite{Kas05,Kas06}.  This is a natural consequence of the more rapid reduction of QP scattering rate than the $N(0)$ reduction, since the number of QPs and the number of QP scatters are both reduced below $T_{c0}$ in the electron-electron scattering. Under a magnetic field, however, $\kappa/T$ begins to decrease below a distinct cusp (dashed arrows) as the temperature is lowered. This is an indication that $N(0)$ decreases below this cusp temperature.  Indeed, according to recent theories \cite{Niv02}, thermal conductivity has no fluctuation correction, in contrast to the resistivity, magnetic susceptibility and specific heat which are subject to the fluctuations.  Therefore, it is natural to consider that the cusp temperature of $\kappa/T$ corresponds to the mean field transition temperature $T_c(H)$.  The decrease of $\kappa/T$ below $T_c(H)$ indicates that $\ell$ remains short under magnetic fields, which shows a clear contrast to the enhanced $\kappa/T$ below $T_c(H)$ in {\it e.g.} CeCoIn$_5$ \cite{Kas05}. 
Further lowering the temperature brings a second anomaly below which $\kappa/T$ becomes bigger than that extrapolated from high temperatures.  This second anomaly is located close to $T_m$ (solid arrows) but far from $T_c(H)$, indicating that the QP scattering is dramatically changed at $\sim T_m$, which will be discussed later.

In Fig.~1, $T_c(H)$ determined by the thermal conductivity is shown by dashed arrows.   It is obvious that $\rho(T)$ shows only a gradual decrease near $T_c(H)$, while a sudden drop occurs at $T_m$ well below $T_c(H)$. We note that the difference between $T_m$ and $T_c(H)$ becomes more pronounced at higher fields and exceeds 20\% of $T_c$ at 7~T [see Fig.~3(f)]. The features of the resistive transition of URu$_2$Si$_2$ bear striking resemblance to that of clean YBa$_2$Cu$_3$O$_7$, in which the sharp drop of the resistivity is observed in a linear scale at the melting transition without sharp anomaly at $T_c(H)$ and the $E$-$J$ characteristics becomes strongly non-Ohmic below $T_m$ \cite{Kwo92}.   Based on these results, we conclude that the melting transition takes place at $T_m$ \cite{mag}.

\begin{figure}[t]
\includegraphics[width=7cm]{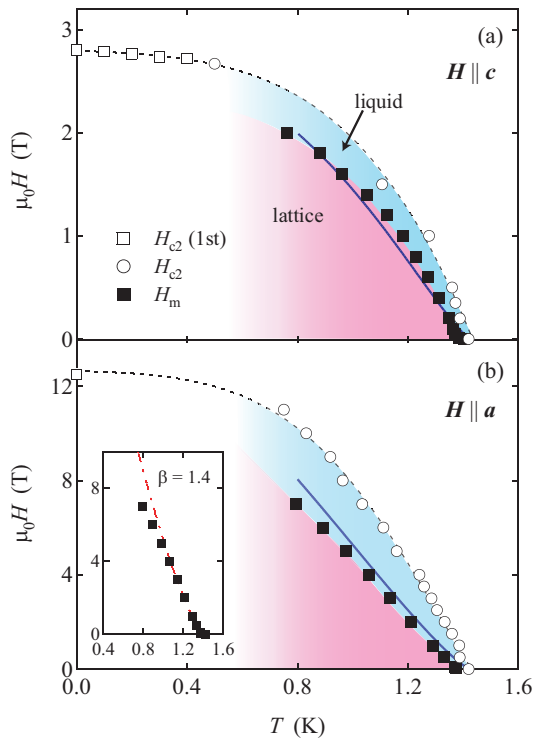}
\caption{(Color online). $H$-$T$ phase diagram of URu$_2$Si$_2$ determined by the present study for (a) {\boldmath $H$} $\parallel c$ and (b) {\boldmath $H$} $\parallel a$. Open symbols represent the mean field $H_{c2}$ lines.  At low temperatures, this line becomes first order (open squares) \cite{Kas07}.  The dashed lines are guides for the eyes.  The solid squares represent the melting transition which is fitted by Eq.~(1) (solid line).  The FL liquid phase occupies a large portion in the $H$-$T$ diagram for both field directions. The inset of (b) shows an expanded view of $H_m(T)$ near $T_{c0}$ for {\boldmath $H$} $\parallel a$.  The dash-dotted line is a fit to $H_m\propto (T_c-T)^{\beta}$ with $\beta=1.4$.  This $\beta$-value is consistent with that in YBa$_2$Cu$_3$O$_7$ \cite{Kwo92}. 
}
\end{figure}

The fundamental parameter which governs the strength of the thermal fluctuations is the Ginzburg parameter, $G_i=[\epsilon k_B T_c/H_{c}(0)^2 \xi_{a}^3]^2/2$, which measures the relative size of the thermal energy $k_BT_c$ and the condensation energy within the coherence volume \cite{Bla94,Larkin}.  Here  $H_{c}=\Phi_0/2 \sqrt{2} \pi \lambda_{a} \xi_{a}$ is the thermodynamic critical field, $\lambda_{a}$ and $\xi_{a}$ are penetration and coherence lengths in the basal plane at $T=0$~K, respectively.  In zero field, the critical region where Gaussian fluctuation breaks down is given by $|T-T_c|/T_c<G_i$.  Such a region is extremely small even in high-$T_c$ cuprates.  However, in magnetic fields sufficiently strong, the superconducting fluctuations acquire an effective one-dimensional (1D) character along the field direction. This reduction of the effective dimensionality increases the importance of fluctuations, resulting in a serious broadening of the resistive transition around $T_c(H)$, particularly in superconductors with large $G_i$ \cite{Ike92}.  Large $G_i$ also leads to the reduction of $T_m$, extending the FL liquid region.    The thermodynamic melting line for 3D system is determined by
\begin{equation}
T_m-T_c(H)=2y\left(\frac{H}{\tilde{H}_{c2}(0)}\right)^{\frac{2}{3}}\sqrt[3]{G_i/\epsilon^2}\ T_c(H)
\end{equation}
with $y\approx -7$ \cite{Larkin}. Here $\tilde{H}_{c2}(0)$ is a linear extrapolation of the initial slope of $H_{c2}(T)$ at $T_{c0}$ to $T\rightarrow 0$~K.

Let us now quantitatively compare  URu$_2$Si$_2$ with other systems. In conventional low-$T_c$ superconductors, $G_i$ ranges from $10^{-11}$ to $10^{-7}$, while in YBa$_2$Cu$_3$O$_7$ $G_i$ is as large as $\sim$10$^{-2}$ \cite{Bla94}. Now, the penetration depth of URu$_2$Si$_2$ is unusually long ($\lambda_{a}\sim$1~$\mu$m according to $\mu$SR \cite{Ama97}), giving rise to a large $G_i\sim3\times10^{-4}$. Such a long penetration depth is a natural consequence of the combination of a small Fermi surface ({\it i.e.} a low carrier density) with a large effective mass. Both these features are directly inferred from de~Haas-van~Alphen measurements \cite{Ohk99} and are confirmed by a host of converging experimental evidence \cite{Beh05}. Thus, $G_i$ is roughly increased by five orders of magnitude, leading to a sizable separation of $T_m$ and $T_c(H)$ over a large portion of the phase diagram, as it is the case in the high-$T_c$ cuprates.

The solid lines in Figs.~4(a) and (b) represent the melting curves obtained from Eq.~(1) with a single fitting parameter $G_i=3.8\times 10^{-4}$.  This value is very close to the above estimate.  These results lead us to conclude that the exceptionally large thermal fluctuations play an important role in URu$_2$Si$_2$ even at subkelvin temperatures.

We here point out several unique features in URu$_2$Si$_2$.  In high-$T_c$ cuprates, the 2D pancake vortices are weakly connected by the Josephson strings for {\boldmath $H$} $\parallel c$ \cite{Cle91},  and the melting transition is accompanied by the strong decomposition of the FLs or ``decoupling" along the $c$-axis \cite{Shi99,Gai00}.    On the other hand, in URu$_2$Si$_2$ where $c$-axis coherence length is much larger than the lattice constant, a 3D melting transition to a line liquid is expected to occur.  Secondly, the transition persists at least up to $\sim0.7H_{c2}(0)$, which is much higher than other systems; Even in very clean YBa$_2$Cu$_3$O$_7$, $H_m(T)$ ceases to increase at $\sim 0.3H_{c2}(0)$, above which the glass transition without sharp resistive drop is observed \cite{Nis00}.  This implies that the present ultraclean URu$_2$Si$_2$ has remarkably few quenched disorder. 
Finally, according to the recent experiments \cite{Kas07}, the first-order transition takes place at $H_{c2}$ below $\sim0.4$~K, owing to the strong Pauli paramagnetism, as shown in Fig.~4.  Consequently, two first-order transition lines with different origins appear to coexist, making the $H$-$T$ phase diagrams of URu$_2$Si$_2$ unprecedented.  Exploring the detailed $H$-$T$ diagram at lower temperatures is therefore intriguing.

The present ultraclean system may also provide important information of the QP transport in the vortex state, which has been a controversial issue \cite{Mat06,Vek99}.  The QP scattering is caused by Andreev scattering on the velocity field associated with the vortices, and a single FL acts as a strong scattering center.  In our URu$_2$Si$_2$, the QP mean free path $\ell$ well exceeds 1~$\mu$m, two orders of magnitude longer than the inter-vortex distance at $\mu_0H$= 1~T.   In a naive picture, such a long $\ell$ would not be influenced by the melting transition. So the observed enhancement of $\kappa/T$ below $T_m$ is highly unusual. Although the full understanding of the unusual change of $\ell$ requires further studies, a possible explanation for this is that below $T_m$ the nearly perfect FL lattice is formed, in which low energy QPs are described by Bloch wave function and are less scattered \cite{Fra00}. It should be noted that the enhancement of $\kappa/T$ in the FL solid state has never been observed even in very clean YBa$_2$Cu$_3$O$_7$ and CeCoIn$_5$ \cite{Kas05}, implying that ultraclean system is required for the formation of the QP Bloch state.

In summary, we provide strong evidence for thermal melting of FL lattice 
to FL liquid at subkelvin temperatures in URu$_2$Si$_2$. The periodic 
FL lattice is suggested to form the QP Bloch state with long mean free path. 
The present results demonstrate that heavy fermion superconductors may provide a new playground to study novel vortex matter physics as well as quasiparticle dynamics in the vortex state of type-II superconductors. 

We thank H. Adachi, G. Blatter, R. Ikeda, T. Kita, N. Kokubo, K. Machida, T. Nishizaki, Z. Te\ifmmode \check{s}\else \v{s}\fi{}anovi\ifmmode \acute{c}\else \'{c}\fi{}, and I. Vekhter for valuable discussions.


\begin{thebibliography}{999}

\item[$^\ast$] Email: okazaki@scphys.kyoto-u.ac.jp
\item[$^\dag$] Email: shibauchi@scphys.kyoto-u.ac.jp 
\item[$^\ddag$] Email: matsuda@scphys.kyoto-u.ac.jp

\bibitem{Pal85}
T.~T.~M. Palstra {\it et al.}, 
Phys. Rev. Lett. {\bf 55}, 2727 (1985).

\bibitem{Amitsuka}
H. Amitsuka {\it et al.}, 
J. Magn. Magn. Mater. {\bf 310}, 214 (2007). 

\bibitem{Ohk99}
H. Ohkuni {\it et al.}, 
Philo. Mag. B {\bf 79}, 1045 (1999).

\bibitem{Kas07}
Y. Kasahara {\it et al.}, 
Phys. Rev. Lett. {\bf 99}, 116402 (2007).

\bibitem{Ada07}
H. Adachi and M. Sigrist, 
arXiv:0710.3110 (2007). 

\bibitem{Beh05}
K. Behnia {\it et al.}, 
Phys. Rev. Lett. {\bf 94}, 156405 (2005).

\bibitem{Kwo92}
W.~K. Kwok {\it et al.}, 
Phys. Rev. Lett. {\bf 69}, 3370 (1992).

\bibitem{Zel95}
E. Zeldov {\it et al.}, 
Nature {\bf 375}, 373 (1995).

\bibitem{Sch97}
A. Schilling {\it et al.}, 
Nature {\bf 382}, 791 (1996).

\bibitem{Wel96}
U. Welp, J.~A. Fendrich, W.~K. Kwok, G.~W. Crabtree, and B.~W. Veal, 
Phys. Rev. Lett. {\bf 76}, 4809 (1996).

\bibitem{Pal00}
In some low-$T_c$ superconductors, disorder driven transition from ordered to disorder glass has been observed. See {\it e.g.} 
Y. Paltiel {\it et al.}, Nature {\bf 403}, 398 (2000).

\bibitem{Nis00}
T. Nishizaki, T. Naito, S. Okayasu, A. Iwase, and N. Kobayashi, 
Phys. Rev. B {\bf 61}, 3649 (2000).

\bibitem{Beh92}
K. Behnia {\it et al.}, 
Physica C {\bf 196}, 57 (1992).

\bibitem{Kas05}
Y. Kasahara {\it et al.}, 
Phys. Rev. B {\bf 72}, 214515 (2005).

\bibitem{Kas06}
Y. Kasahara {\it et al.}, 
Phys. Rev. Lett. {\bf 96}, 247004 (2006).

\bibitem{Niv02}
D.~R. Niven and R.~A. Smith, 
Phys. Rev. B {\bf 66}, 214505 (2002);
S. Vishveshwara and M.~P.~A. Fisher, 
Phys. Rev. B {\bf 64}, 134507 (2001).

\bibitem{mag}
To obtain thermodynamic evidence of the first-order transtion, we have also tried the local magnetization measurements by using a small Hall sensor of HgCdTe with an active area of $100\times100~\mu$m$^2$, which is placed on the surface of URu$_2$Si$_2$. However, the jump of the magnetization 
is not resolved within experimental resolution.  The jump $\Delta B$ associated with the first-order melting transition is given as, $\Delta B$\,[G]~$\approx (1.5\times 10^{-6}\epsilon) T_m$\,[K]$\times(B_m$\,[G]$)^{1/2}$, where $\epsilon=H_{c2}^a/H_{c2}^c$ is the anisotropy parameter \cite{Dog98}. Using $\epsilon\approx4$ for URu$_2$Si$_2$,  $\Delta B$ is estimated to be $1.8\times10^{-4}$~G at $T_m=1.4$~K and $B_m=500$~G, which is indeed found to be much smaller than our resolution $\sim10^{-2}$~G. 

\bibitem{Dog98}
M.~J.~W. Dodgson, V.~B. Geshkenbein, H. Nordborg, and G. Blatter, 
Phys. Rev. Lett. {\bf 80}, 837 (1998).

\bibitem{Bla94}
G. Blatter, 
M.~V. Feigel'man, V.~B. Geshkenbein, A.~I. Larkin, and V.~M. Vinokur, 
Rev. Mod. Phys. {\bf 66}, 1125 (1994). 

\bibitem{Larkin}
A. Larkin, and A. Varlamov, {\it Theory of fluctuations in superconductors} 
(Oxford Univ. Press, Oxford, 2005).

\bibitem{Ike92}
R. Ikeda, T. Ohmi, and T. Tsuneto, 
J. Phys. Soc. Jpn. {\bf 58}, 1377 (1989).

\bibitem{Ama97}
A. Amato, 
Rev. Mod. Phys. {\bf 69}, 1119 (1997). 

\bibitem{Cle91}
J.~R. Clem, 
Phys. Rev. B {\bf 43}, 7837 (1991).

\bibitem{Shi99}
T. Shibauchi {\it et al.}, 
Phys. Rev. Lett. {\bf 83}, 1010 (1999). 

\bibitem{Gai00}
M.~B. Gaifullin, Y. Matsuda, N. Chikumoto, J. Shimoyama, and K. Kishio, 
Phys. Rev. Lett. {\bf 84}, 2945 (2000).

\bibitem{Mat06}  
Y. Matsuda, K. Izawa, and I. Vekhter, 
J. Phys.: Condens. Matter {\bf 18}, R705 (2006).

\bibitem{Vek99} 
I. Vekhter and A. Houghton, 
Phys. Rev. Lett. {\bf 83}, 4626 (1999).

\bibitem{Fra00}
M. Franz and Z. Tesanovic, 
Phys. Rev. Lett. {\bf 84}, 554 (2000); 
K. Yasui and T. Kita, 
Phys. Rev. Lett. {\bf 83}, 4168 (1999);
M.~R. Norman, A.~H. MacDonald, and H. Akera, 
Phys. Rev. B {\bf 51}, 5927 (1995).
 

\end{thebibliography}
\end{document}